\documentclass[floatfix,aps,prl,twocolumn,superscriptaddress]{revtex4}

\usepackage[dvips]{graphicx,color}      
\usepackage{amsmath,amssymb}
\begin{document}

\newcommand{\avg}[1]{\langle #1 \rangle}
\newcommand{\Neff}{N_{\rm eff}}
\newcommand{\Vcav}{V_{\rm cav}}
\newcommand{\Var}{{\rm Var}}


\title{Fast cavity-enhanced atom detection with low noise and high fidelity}

\author{J. Goldwin}
\affiliation{Centre for Cold Matter, Imperial College, Prince Consort Road, London SW7 2BW, United Kingdom}
\affiliation{School of Physics and Astronomy, University of Birmingham, Edgbaston, Birmingham B15 2TT, United Kingdom}
\author{M. Trupke}
\affiliation{Centre for Cold Matter, Imperial College, Prince Consort Road, London SW7 2BW, United Kingdom}
\affiliation{Vienna Center for Quantum Science and Technology, Atominstitut, TU Wien, 1020 Vienna, Austria}
\author{J. Kenner}
\author{A. Ratnapala}
\author{E. A. Hinds}
\affiliation{Centre for Cold Matter, Imperial College, Prince Consort Road, London SW7 2BW, United Kingdom}


\begin{abstract}
Cavity quantum electrodynamics describes the fundamental interactions between light and matter, and how they can be controlled by shaping the local environment.  For example, optical microcavities allow high-efficiency detection and manipulation of single atoms.  In this regime fluctuations of atom number are on the order of the mean number, which can lead to signal fluctuations in excess of the noise on the incident probe field.  Conversely, we demonstrate that nonlinearities and multi-atom statistics can together serve to suppress the effects of atomic fluctuations when making local density measurements on clouds of cold atoms.  We measure atom densities below 1 per cavity mode volume near the photon shot-noise limit.  This is in direct contrast to previous experiments where fluctuations in atom number contribute significantly to the noise.  Atom detection is shown to be fast and efficient, reaching fidelities in excess of $97\%$ after $10\,\mu$s and $99.9\%$ after $30\,\mu$s.
\end{abstract}

\maketitle

\section{Introduction}

High finesse optical resonators can improve the sensitivity of atom detection by increasing the lifetime of photons and confining them to a small volume \cite{Ye03}.  Long photon lifetime, controlled by cavity length and mirror reflectivity, increases the effective optical thickness of an intra-cavity sample by a factor on the order of the finesse $\mathcal{F}\gg 1$.  Small mode volume, which depends only on the geometry of the resonator, increases the energy density per photon and therefore the Einstein coefficients describing transition rates.  Thus the spontaneous emission rate of an atom is increased by coupling it to a resonant cavity \cite{Pur46}.  Importantly, all the additional photons are emitted into the cavity mode, making it possible to detect fluorescence even at very low atom density. For sufficiently small mode volumes, a single cavity photon becomes intense enough to saturate the atomic transition.  In this regime vacuum fluctuations modify the spectral properties of the coupled atom-cavity system \cite{Jay63} in such a way as to allow detection at the single-atom level \cite{Tho92,Chi96,Mun00,Tru07a}.

Recently there has been growing interest in cold atom experiments with atomic density distributions extending throughout or beyond the range of the cavity field \cite{Bre07,Col07,Gup07,Tep08,Sch10}.  For multiple atoms, the radiative behaviour can be coherent \cite{Dic54,Tav68}.  Although the gas may be dilute, the common coupling to the electromagnetic field produces effective long-range interactions between the atoms that can lead to self-organisation \cite{Cha03,Bla03} and collective motion \cite{Nag03}, as well as super-radiant Rayleigh scattering and collective atomic recoil lasing \cite{Sla07}.  Recently experimenters have exploited these effects to realise a quantum phase transition from a Bose-Einstein condensate to a supersolid \cite{Bau10}.

A central parameter in describing cavity-enhanced detection is the dimensionless single-atom cooperativity \cite{Hor03}, \mbox{$C_1=g^2/(2\kappa\gamma)$}, where $2\,g$ is the single-photon Rabi frequency at the peak of the cavity intensity distribution, $2\,\kappa$ is the cavity linewidth (full-width at half maximum), and $2\,\gamma$ is the natural atomic linewidth.  The cooperativity determines both the effect of a single atom on the cavity spectrum, and the rate of fluorescence into the cavity.

In the case of multiple atoms, the cooperativity is generalized by defining $C_N=C_1\,\Neff$, where the effective atom number is \cite{Car99}
\begin{eqnarray}\label{eq:Neff}
\Neff &=& \int\limits_0^L\!\int\limits_{-\infty}^\infty\!\int\limits_{-\infty}^\infty\! |\chi({\mathbf r})|^2\varrho(\mathbf{r})\,{\rm d}^3{\mathbf r}
\end{eqnarray}
with $\varrho({\mathbf r})$ being the atomic density, $L$ the cavity length, and $\chi(\mathbf{r})=\sin(2\pi\,z/\lambda)\exp[-(x^2+y^2)/w^2]$ the cavity field mode function ($\lambda$ is the wavelength).  It is important to note that $\Neff$ is a random variable, generally distinct from its mean value $\avg{\Neff}$. When the atom cloud is much larger than the cavity mode volume \mbox{$\Vcav=\pi w^2L/4$}, the mean atom density is approximately uniform over the interaction region, and $\avg{\Neff}\approx \varrho(0)\,\Vcav$. At low densities, single-atom physics dominates, while at higher densities multi-atom effects become important \cite{Car99}.  Here we perform local density measurements on large dilute clouds of atoms in the crossover regime, paying particular attention to signal fluctuations. We show that even at densities on the order of one atom per cavity mode volume, the effects of atomic shot noise are heavily suppressed.  We then compare our results with state-of-the-art experiments on single trapped atoms, demonstrating a fast detection time and high fidelity.

\section{Results}
\subsection{Optical Noise Suppression}

\begin{figure}
\includegraphics[scale=0.4]{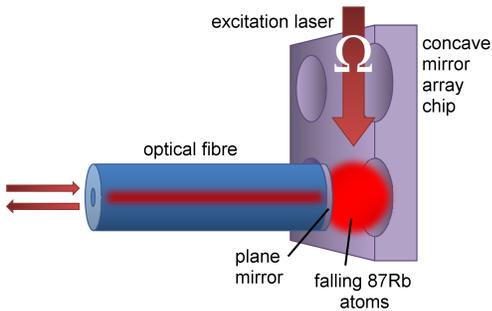}
\caption{\label{fig:expt}Schematic of experiment (not to scale).  Clouds of $^{87}$Rb atoms are laser cooled and dropped through a high-finesse optical microcavity.  In the experiment, the atomic density is approximately Gaussian with a width three orders of magnitude larger than the cavity mode waist, allowing us to make a local density approximation for the atoms.  The system is probed either by monitoring the cavity reflection, with $J_{\rm in}$ incident photons per second, or by observing fluorescence into the cavity mode induced by a laser beam with Rabi frequency $\Omega$ (and $J_{\rm in}=0$).  In either case the output stream of $J_{\rm out}$ photons per second is detected using a single photon counting module.}
\end{figure}

Our apparatus, shown schematically in Fig.~\ref{fig:expt}, has been described in detail in refs.~\cite{Tru07a,Tru05} (see also Methods).
We detect atoms either by (\emph{i}) measuring changes in the intensity of a probe beam reflected from the cavity; or (\emph{ii}) detecting fluorescence when exciting the atoms uniformly with a laser beam propagating transverse to the cavity axis.  We refer to these simply as reflection and fluorescence measurements, respectively. If atoms are present and the cavity and lasers are resonant with the free-space atomic transition, then the steady-state rate of photons travelling from the cavity to the detector is \cite{Tru07b}
\begin{eqnarray}\label{eq:signals}
J_{\rm out} &=& \left\{
\begin{array}{ccl}
\displaystyle J_{\rm in}\left(\frac{b+2\,C_N}{1+2\,C_N}\right)^2 &\;,\quad& {\rm reflection}\\~\\
\displaystyle 2\,C_N^\prime\,\gamma\,\xi\;\frac{s}{(1+2\,C_N^\prime)^2 + s} &\;,\quad& {\rm fluorescence}
\end{array}\right.
\end{eqnarray}
where $J_{\rm in}$ is the number of incident probe photons per second and $b^2$ characterizes the reflection fringe contrast in the absence of atoms.  The cooperativities for reflection ($C_N$) and fluorescence ($C_N^\prime$) are not generally the same since they depend on the polarisation of the probe light and the excitation light respectively~\cite{Ken10} (see Methods). In fluorescence, $s=\tfrac{1}{2}(\Omega/\gamma)^2$ is the free-space saturation parameter for excitation driven at a Rabi frequency $\Omega$, while $\xi$ is the probability for an intracavity photon to pass from the cavity into the fibre.  Finally we have used the facts that $(g/\kappa)^4\ll 1$, and that the atomic excited state fraction is small in our reflection measurements.  For this work \mbox{$g/(2\pi)= 98.4(1.6)\,$MHz}, \mbox{$\kappa/(2\pi)= 5200(100)\,$MHz}, and $\gamma/(2\pi)= 3\,$MHz, giving $C_1=0.307(11)$. 

It is important to note that $J_{\rm out}$ as described by Eq.~(\ref{eq:signals}) is only linear in atom density for small values of the cooperativity; for reflection measurements $J_{\rm out}$ saturates with increasing $C_N$, while for fluorescence $J_{\rm out}$ reaches a maximum when $C_N^\prime=\tfrac{1}{2}(1+s)^{1/2}$ and then rolls over and vanishes.  We will show that this allows us to operate in a regime where we remain sensitive to variations in mean atomic density while damping out the effects of large instantaneous fluctuations.

\begin{figure}
\includegraphics[scale=0.5]{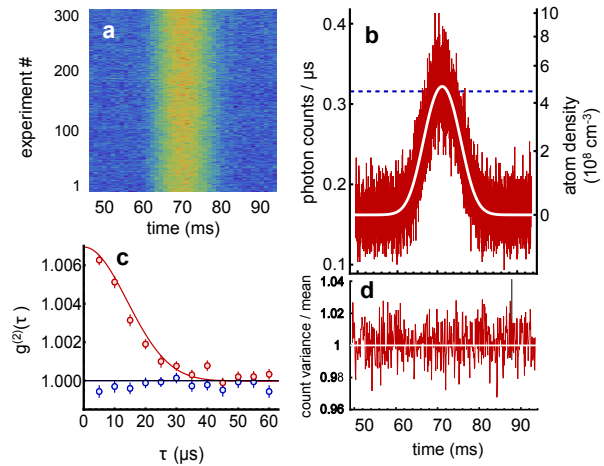}
\caption{\label{fig:Reflection}Reflection measurements. (a) Detected photon counts for 300 identical experiments.  The atoms are released at $39.5$\,ms.  Counts increase from blue to red.  Data were taken with $2\,\mu$s resolution, and the image was then re-binned to 1 ms.  (b) The data in red are averages over the 300 drops shown in (a), while the white curve is a fit to Eq.~(\ref{eq:signals}) assuming a Gaussian dependence of $C_N$ on time; the dashed blue line gives the value expected from a single atom maximally coupled to the cavity mode.  (c) Second order correlation $g^{(2)}(\tau)$ as a function of time delay $\tau$.  Red points are from data with $\avg{\Neff}=0.225(17)$, and blue points are taken without atoms; error bars show the standard error of the mean at each delay time from 50 trials.  The solid curves are the theoretical expectations.  (d) Ratio of ensemble variance to mean versus time.  The red curve is calculated from the raw data in (a) for each $2\,\mu$s time bin, and then a $100\,\mu$s running average is applied to smooth the result; the white line is the photon shot-noise level.}
\end{figure}

Figure \ref{fig:Reflection}(a) shows the results of repeated reflection measurements of clouds falling through the cavity.  A circularly-polarized probe drives the atomic cycling transition, maximizing the atom-field coupling strength. At early and late times, there are no atoms in the cavity, so the reflected light is at its minimum value, determined by the incident probe power and the empty cavity fringe contrast. The reflected intensity rises when there are atoms in the cavity. These experimental runs are averaged in Figure \ref{fig:Reflection}(b), which also shows a fit to the evolution expected from Eq.~(\ref{eq:signals}) with a peak $\avg{\Neff}=1.06(4)$, corresponding to only $4.9(2)\times10^8\,$atoms $\rm{cm}^{-3}$. For reference, the dashed line shows the expected reflection with a single atom maximally coupled to the cavity mode. Note that the $10\,$ms width of the curve reflects the size of the cloud, which is determined by its temperature. By contrast, the typical transit time for a single atom passing through the width of the cavity mode is $\sim 14\,\mu$s, so the cloud is very large compared with the extent of the cavity field.  Individual transits are revealed in Fig.~\ref{fig:Reflection}(c), where we show the measured second order (intensity) correlation, given by
\begin{eqnarray}
\label{eq:g2}
g^{(2)}(\tau) = \frac{\overline{k(t)k(t+\tau)}}{\overline{k(t)}^{\,2}}
\end{eqnarray}
where $k$ is the number of photons counted during a time window centred on $t$, and $\tau$ is the relative delay between windows.  Overbars denote an average over $t$ using a $500\,\mu$s segment of data, throughout which $\Neff$ is approximately constant. The solid red curve shows the expectation for a single atom crossing the cavity according to Eq.~(\ref{eq:signals}).  The only free parameter is the amplitude of the peak, which accounts for having more than one atom pass through the cavity during the trace, but with not all atoms optimally coupled.  In Fig.~\ref{fig:Reflection}(d) we plot as a function of time the variance of the photon counts divided by the mean, evaluated over the 300 repetitions of the experiment. Although one might have expected to see an increased variance with the arrival of the atom cloud, there is in fact no sign of such an increase. We return to this point below, when we see similarly low noise levels in our fluorescence measurements.

\begin{figure}
\includegraphics[scale=0.45]{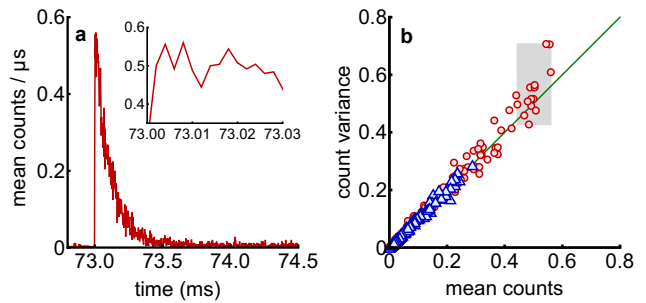}
\caption{\label{fig:Fluorescence}Fluorescence measurements.  (a) Fluorescence pulse, averaged over 250 drops. The exciting laser is pulsed on after the peak atomic density has passed through the cavity. At the start of the pulse, shown in detail in the inset, $\avg{\Neff}=1.24(5)$. This is slightly larger than in Fig.~\ref{fig:Reflection} due to a higher atom number in the initial MOT. (b) Variance of fluorescence counts as a function of mean.  Red circles are from the data used in (a), and blue triangles are from a set where the drive beam is pulsed on at a later time in the drop, with initial $\avg{\Neff}=0.50(2)$.  The green line is the photon shot-noise limit.  The grey box corresponds to the inset in (a).}
\end{figure}

Figure \ref{fig:Fluorescence}(a) shows the fluorescence signal.  As the cloud falls through the cavity we switch on a resonant excitation beam whose (downward) propagation direction and polarisation are both perpendicular to the cavity axis. The photon count rate immediately jumps to a high level as a result of the laser-induced fluorescence. Independent reflection measurements determine that the initial $\avg{\Neff}=1.24(5)$. Although the atom number is nearly constant over several ms during the reflection measurements, the signal here decays roughly exponentially with a time constant of order $100\,\mu$s. This is because the atoms are heated and pushed out of the cavity by the excitation light~\cite{Ken10}, which is much more intense than the probe light used in reflection measurements. In Fig.~\ref{fig:Fluorescence}(b) we plot how the variance of the fluorescence count over 250 repetitions varies with the mean number of counts. Once again, we see that the fluctuations are very near the photon shot-noise limit, which is indicated by the solid line.

We reiterate that these results are in direct contrast to similar experiments operating in the linear regime, \emph{i.e.}, when the photon counts are strictly proportional to atom number or density.  For example, we can compare our results in Fig.~\ref{fig:Fluorescence}(b) with Fig.~2(b) of ref.~\cite{Wil06}, which used micro-optics to detect atoms without a cavity.  They observed that the ratio of variance to mean doubled in the presence of atoms. To calculate the noise level for fluorescence detection in the linear regime, one can apply Mandel's theory as described in ref.~\cite{Tep06}. Atomic motion is negligible over a single $1\,\mu$s time bin, so we consider each bin to have a fixed number of atoms $N$, producing a Poissonian photon count $k$ with a mean of $\alpha$ photons per atom (the background count is negligible compared to $\alpha$). Since $N$ fluctuates over repeated experiments, the photon counts obey $\Var(k)/\avg{k}=1+\alpha\Var(N)/\avg{N}$. If atoms are positioned randomly with a uniform probability distribution, the number of atoms in a given volume follows a Poisson distribution and $\Var(k)/\avg{k} = 1+\alpha$, independent of $\avg{N}$. In our experiment, the yield of fluorescence photons for one (hypothetical, maximally coupled) atom is $0.42(2)$ in $1\,\mu$s.  Clearly this theory does not describe our experiment, whose measured value of $[\Var(k)/\avg{k}-1] = 0.09(3)$ is much smaller.

Our measurements require a different analysis, as the assumption of linearity is violated in Eq.~(\ref{eq:signals}) and the variable $\Neff$ should be considered rather than $N$. We proceed as follows.  Our experiment operates with mean intracavity photon number $\avg{n}\equiv\avg{a^\dagger a}\ll 1$, and $\kappa^{-1}\ll g^{-1}$, meaning excitations of the cavity field result in immediate emission of photons rather than reabsorption by atoms.  Over a sufficiently short time $T$, the probability of emitting a photon is just $2\kappa\avg{a^\dagger a}T$.  Then we have
\begin{eqnarray}
\label{eq:ourVark}
\frac{\Var(k)}{\avg{k}} = 1 + 2\kappa T\xi\epsilon\, \frac{\Var(\!(n)\!)}{\langle\!\langle n\rangle\!\rangle}
\end{eqnarray}
where $\epsilon$ is the total collection efficiency from fibre to detector, and $T$ is assumed to be much longer than any correlation times for the fluorescence (\emph{e.g.} $1/\gamma$, $\kappa/g^2$). Double brackets denote statistics taken over the conditional distribution of $n$ given a fixed $\Neff$, and the distribution of $\Neff$ itself. We obtain $n$ via the master equation for the density matrix, and use the results of ref.~\cite{Car99} for the probability density for $\Neff$ (see Methods).  The value of $\avg{k}$ obtained in this way is almost identical to what one obtains by simply setting $C_N=C_1\avg{\Neff}$ in Eq.~(\ref{eq:signals}). For the fluctuations we obtain $\Var(k)/\avg{k}=1.095(8)$, in excellent agreement with our observed value. A similar treatment of the reflection measurements in Fig.~(\ref{fig:Reflection}) gives $\Var(k)/\avg{k}=1.005(2)$, consistent with the value $1.002(4)$ from the data.  Note that for the two types of measurement the ratio $\Var(\Neff)/\avg{\Neff}=3/8$ is the same, but the nonlinearity of $J_{\rm out}$ is quite different. Calculations for both types of measurement, with the same $\avg{\Neff}$ and $J_{\rm in}$ adjusted to have equal numbers of signal photons, show that the noise suppression is still much stronger for reflection. This stems from the saturation of the reflection signal at large instantaneous $\Neff$ versus the roll-over of the fluorescence.  For both measurements we conclude that the statistics of $\Neff$ and the nonlinearity of the interactions are jointly responsible for the strong optical noise suppression that we observe.

\subsection{Discrete Detection}

Our measurements do not involve trapping single atoms within the cavity mode.  However we have shown that our signal fluctuations are near the photon shot noise limit, effectively allowing us to neglect fluctuations in $\Neff$. This allows a direct comparison between our measurements on falling clouds and experiments where noise in the atom number is inherently absent due to preparation of single trapped atoms.  In this context we discuss the discrete detection problem, for example distinguishing between hyperfine ground states.  Since the detection linewidth is three orders of magnitude smaller than the level splitting, the $|F=1\rangle$ ground state is effectively dark in our system \cite{Ken10}. When atom number fluctuations are suppressed, discrete detection with $\avg{\Neff}=1$ is thus equivalent to the problem of determining whether a single trapped atom fluoresces or not, which is relevant for quantum information processing \cite{Boc10,Geh10}.  We therefore take detection of $\avg{\Neff}=1$ as the benchmark for comparison with other experiments. From our fluorescence measurements at $\avg{\Neff}=1.24(5)$, we extrapolate a mean photon count rate at $\avg{\Neff}=1$ of $S_1=420(20)\,$ms$^{-1}$. Table \ref{tab:comparison} shows that this is high in comparison with other atom detection experiments. Following ref.~\cite{Tep06}, we could define the single-atom efficiency of the detector as $\eta=1-\exp(-S_1T)$. This is the probability of counting $\ge 1$ photon during the measurement time $T$, when an atom is present and assuming Poissonian photon counts with negligible background. This rises rapidly with our high count rate, reaching $98.5(3)\%$ in only $10\,\mu$s.

\begin{table}
\begin{tabular}{ccccccc}\hline\hline
Ref.		         & $S_1$  				& $B$				& $F_{1\rm{max}},T_{1\rm{max}}$		& $F_{2\rm{max}}$		\\ \hline
\cite{Tep06}      & $5.6$					& $0.28$			& $90.9,544$     								& $97.5$					\\
\cite{Wil06}      & $36$            	& $0.311$     	& $97.6,132$      							& $99.8$					\\
\cite{Ter09}      & $54.5$          	& $2.18$   		& $92.2,60$      								& $98.1$					\\
\cite{Koh09}      & $0.13$          	& $0.05$			& $72.1,9853$      							& $80.5$					\\
\cite{Boc10}		& $94$					& $0.05$			& $99.773,80$									& $99.99982$			\\
\cite{Geh10}		& $190$					& $1.4$			& $97.87,26$									& $99.85$				\\
This work         & $\;\;420(20)$   	& $3.84(6)$		& $97.46(13),11.2(4)$ 						& $99.79(2)$			\\ \hline\hline
\end{tabular}\caption{\label{tab:comparison}Comparison with other experiments.  Rates are in cts/ms, fidelities in percent, and $T_{1\rm{max}}$ in $\mu$s. Note that $T_{K\rm{max}}=K\,T_{1\rm{max}}$ for $p=1/2$.  References \cite{Tep06,Ter09,Boc10,Geh10} use cavities, while refs.~\cite{Wil06,Koh09} use optical waveguides without cavities.  References \cite{Boc10,Geh10} describe non-destructive detection.}
\end{table}

\begin{figure}
\includegraphics[scale=0.42]{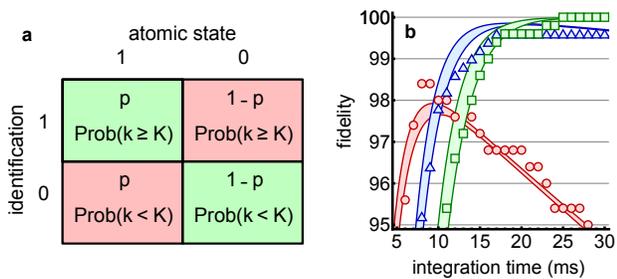}
\caption{\label{fig:Fidelity}
Detection fidelity.  (a) Calculating the fidelity.  Given two atomic states there are four possible outcomes of the experiment.  Columns (rows) correspond to the actual (identified) state.  In each box the upper quantity is the probability to have the state, and the lower is the probability to make the identification; the total probability for the corresponding outcome is the product of the two.  Correct identifications are in green (diagonal), and their sum equals the fidelity as given by Eq.~(\ref{eq:FK}) in the text. (b) Detection fidelity $F_K(T)$ for $K=1$, 2, and 3 counts (red $\circ$, blue $\vartriangle$, and green $\square$).  Points are from 500 measurements with $p=1/2$ and curves show Eq.~(\ref{eq:FK}), assuming Poisson distributions with mean signal and background count rates obtained from the data at the beginning of the pulse (the width of the curves reflects the statistical uncertainties in these rates).  The steps in the data are in units of the minimum resolution of $0.2\%$ for 500 trials.}
\end{figure}

For most applications, however, it is not enough to detect the bright state (logical 1) efficiently; the detector must also be able to identify the dark state (logical 0) correctly. A more useful figure of merit is thus the fidelity, which is the probability of a correct measurement result. Let us take the detection of $\ge K$ photons as indicating logical 1, and $<K$ as logical 0. Then for Poissonian distributions the single-photon fidelity is $F_{K=1}=(1-p)\,e^{-BT}+p\,[1-e^{-(S+B)T}]$, where $B$ is the background photon counting rate and $p$ is the probability that the state being measured is logical 1. The first (second) term is the probability of having logical 0 (1) and identifying it correctly. The four logical possibilities are shown schematically in Fig.~\ref{fig:Fidelity}(a).  The red curve in Fig.~\ref{fig:Fidelity}(b) shows the expected value of $F_{K=1}$ in our experiment over a data set for which $\avg{\Neff}=1.24$ and $p=1/2$. The fidelity rises quickly as the detection of logical 1 becomes increasingly successful, but eventually falls due to false positives from the background. Superimposed on this curve are our measured values of the fidelity versus detection time, which agree well with our expectations. In general the maximum fidelity $F_{1\rm{max}}$ increases with $S/B$, reaching its peak at a time $T_{1\rm{max}}$ proportional to $1/S$ for fixed $S/B$.  

Table~\ref{tab:comparison} compares our values of $F_{1\rm{max}}$ and $T_{1\rm{max}}$ with those for other atom detection experiments.  The highest fidelity by far is that of ref.~\cite{Boc10}, while our high signal rates result in the shortest detection time. It is worth noting that the measurements in refs.~\cite{Boc10,Geh10} are non-destructive, whereas the rest are carried out on resonance. Lossless fluorescence detection of single trapped atoms in free space has been observed with $95\%$ $(98.6\%)$ accuracy in $0.3$ ms ($1.5$ ms) \cite{Gib11,Fuh11}.  Additionally, recent refinements to our cavity manufacturing process have increased the finesse by two orders of magnitude \cite{Bie10}.  This suggests the possibility of single-atom strong coupling with $g>(\kappa,\gamma)$ and $C_1$ in the hundreds, allowing non-destructive measurements in our system as well.

A simple way to improve the fidelity is to increase the detection threshold $K$. This leads to the general result
\begin{eqnarray}\label{eq:FK}
F_K = (1-p)\,\frac{\Gamma[K,BT]}{(K-1)!} + p\left[1-\frac{\Gamma[K,(B+S)T]}{(K-1)!}\right]
\end{eqnarray}
where $\Gamma [K,a]$ is the incomplete gamma function. These fidelities are plotted in Fig.~\ref{fig:Fidelity} versus measurement time $T$ for the cases of $K=2$ and 3. They peak at $99.79\%$ and $99.98\%$ when $T=22.4\,\mu$s and $33.6\,\mu$s respectively. The data points again show that our measurements are consistent with expectations. Similar methods were exploited in \cite{Ter09}, where two photons were required within a short time window in order to register a logical 1 result. With a $1\,\mu$s detection window they found that $99.719(6)\,\%$ of observed 2-photon coincidences were due to atoms, and described this percentage as the fidelity. In that experiment however, there was only a $0.2\,\%$ chance that the logical 1 state would produce a 2-photon count in the detection window. Thus, although the detection confidence was high, the efficiency was low, resulting in a low fidelity in the usual sense that we adopt here.

\section{Discussion}

We have characterized a cavity-enhanced atom detector with low noise and high spatial resolution (set by the small cavity mode waist).  We have shown that the nonlinear, multi-atom nature of the interactions results in a strong suppression of signal noise due to atomic fluctuations.  Our detector is fast and efficient, and suitable for detecting dilute samples below the level of a single atom per mode volume. Although we have focused here on measurements of low atomic densities, the dynamic range can be extended upwards simply by detuning the cavity and/or the probe field.

We envision a variety of applications for making local density measurements on cold atom clouds and quantum gases.  For example, small impurities can be detected for studies of Fermi polaron physics \cite{Sch09} and quantum transport \cite{Pal09}.  Cavity-enhanced detection also allows a greater collection efficiency for scattered photons than in conventional high-numerical-aperture (NA) optical systems \cite{Hor03}.  The maximum fraction of photons which can be captured in such systems is approximately ${\rm NA}^2/4$, which even for the best available lenses is an order of magnitude smaller than the fraction $2C_N/(1+2C_N)$ which can be captured by a cavity with $C_N\sim1$.  This could improve the speed and efficiency of atom trap trace analysis, where laser-induced fluorescence is used to detect radioactive atoms for dating environmental samples over time scales not accessible with $^{14}$C \cite{Che99}.  Finally, the compatibility of our detector with atom chips makes it attractive for studying quantum gases in the Tonks-Girardeau regime \cite{Rei04}.  Producing one-dimensional gases requires trapping potentials of extremely high aspect ratio, as are typical with atom chips, and strong interactions require low densities which can be detected locally very quickly with our cavity.


\section{Methods}
\subsection{Experiment}
We work with $^{87}$Rb, near the D$_2$ spectral lines at $\lambda=780\,$nm. Our optical microcavity is formed between the end of a single-mode optical fibre and a spherical surface microfabricated in silicon, both being coated with multilayer dielectric mirrors. The resulting plano-concave cavity mode has a length of $L=139(1)\,\mu$m and a waist whose $e^{-1}$ field radius is $w=4.46(7)\,\mu$m. To our knowledge the only Fabry-Perot cavity with a smaller mode waist is the all-fibre design of ref.~\cite{Col07} ($w=3.9\,\mu$m).  Since $C_1\propto \mathcal{F}/w^2$, a small waist makes it possible to detect single atoms using a cavity of relatively modest finesse. This relaxes the usual need for very high mirror quality and reduces the sensitivity to noise in the cavity length.  As stated in the main text, \mbox{$g/(2\pi)= 98.4(1.6)\,$MHz}, \mbox{$\kappa/(2\pi)= 5200(100)\,$MHz}, and $\gamma/(2\pi)= 3\,$MHz, giving $C_1=0.307(11)$. We begin each experimental sequence by cooling and trapping $\sim 2\times 10^7$ $^{87}$Rb atoms in a magneto-optical trap formed above a mirror \cite{Rei99}, followed by sub-Doppler cooling to $16\,\mu$K in optical molasses.  We then release the atoms, which fall through a hole in the mirror and pass through a cavity mounted immediately below.
  
\subsection{Master Equation}
The Hamiltonian describing our system is ($\hbar=1$)
\begin{eqnarray}
\label{eq:H}
H = -i\eta(a-a^\dagger)-i\sum_{j=1}^N\left[\left(g_ja+\frac{1}{2}\Omega_j\right)\sigma_j^\dagger-{\rm H.c.}\right]
\end{eqnarray}
where we have used the rotating wave approximation and neglected atomic centre-of-mass motion.  The operator $a$ annihilates a cavity photon and $\sigma_j=|g\rangle\langle e|$ lowers the $j^{\rm th}$ atom from the excited state $|e\rangle$ to ground state $|g\rangle$, and the pump strength $\eta\propto J_{\rm in }^{1/2}$ \cite{Tru07b}.  We have expressed $H$ in a frame rotating with the angular frequency of the laser, and used the fact that the cavity and fields are on resonance with the free-space $|g\rangle\to |e\rangle$ transition.  The system evolves in time according to the master equation for the density matrix $\rho$
\begin{eqnarray}
\label{eq:master}
\frac{\rm d}{{\rm d}t}\,\rho = -i[H,\rho] - \sum_{j=0}^N\mathcal{D}[A_j]\,\rho
\end{eqnarray}
where $\mathcal{D}[A_j]\,\rho=\tfrac{1}{2}(A_j^\dagger A_j\rho + \rho A_j^\dagger A_j)-A_j\rho A_j^\dagger$, $A_0\equiv(2\kappa)^{1/2}a$, and $A_{j>0}\equiv(2\gamma)^{1/2}\sigma_j$.

For reflection measurements, under the assumption $\avg{a^\dagger a}\ll 1$ and with $\avg{\sigma_j^\dagger\sigma_j}\ll 1$ one obtains a coherent intra-cavity field with amplitude $|\alpha|=\avg{a^\dagger a}^{1/2}$ obeying
\begin{eqnarray}
\label{eq:alphaR}
\alpha = \frac{\eta}{\kappa}\,\frac{1}{1+2C_N}
\end{eqnarray}
The total field reflected from the cavity comprises a component reflected immediately from the input/ouput mirror which interferes with the fraction of the intra-cavity field (\ref{eq:alphaR}) being transmitted back out.  This leads to Eq.~(\ref{eq:signals}) for reflection.

For fluorescence we ignore correlations between atoms, which are distributed randomly, and consider a single atom with coupling $g\,\Neff^{1/2}$ experiencing a Rabi frequency $\Omega$ from the driving laser. The excited state population is
\begin{eqnarray}
\label{eq:Pee}
\left\langle\sigma^\dagger\sigma\right\rangle = \frac{1}{2}\,\frac{\left|\Omega\right|^2/2}{\gamma^2_{\rm tot}+\left|\Omega\right|^2/2}
\end{eqnarray}
where $2\gamma_{\rm tot}$ is the total radiation rate, and we have assumed that the external field is much stronger than the cavity field $|\Omega|\gg 2|g|\,n^{1/2}$.  From the Purcell effect $\gamma_{\rm tot}=(1+2C^\prime_N)\,\gamma$, with a fraction $2C^\prime_N/(1+2C^\prime_N)$ going into the cavity mode \cite{Pur46}.  With the assumptions and definitions in the text, we recover Eq.~(\ref{eq:signals}). To determine $C_N^\prime$ we solved the master equation for a toy model including all 12 of the Zeeman substates of the $F=2$ ground and $F^\prime=3$ excited states, but neglecting the cavity.  The equilibrium excited-state populations were determined and the corresponding total decay rate of $\sigma^\pm$ transitions was calculated (the quantisation axis was taken along the cavity axis, so $\pi$ transitions did not contribute).  The calculated dependence of $J_{\rm out}$ on the polarisation of the drive laser agreed well with experimental results \cite{Ken10}.  The ratio $C^\prime_N/C_N$ has a weak dependence on $s$. In this work $C^\prime_N/C_N=0.53(2)$.  The validity of all our results and conclusions were supported by direct numerical solution of Eqs.~(\ref{eq:H}) and (\ref{eq:master}), as well as quantum jump simulations \cite{Dal92,Gar92,Car93}.  For our parameters the intracavity field is indistinguishable from a coherent state for any fixed arrangement of atoms and either type of detection.

\subsection{Probability Density for $\Neff$}
Carmichael and Sanders derived an expression for the probability density $P(G)\,{\rm d}G$ for the collective dipole $G\equiv\Neff^{1/2}$ in ref.~\cite{Car99}, focusing on the case of travelling-wave cavities.  The distribution depends on $\avg{\Neff}$ and must generally be obtained numerically.  Taking into account our standing wave geometry, and in the limit $\avg{\Neff}\gg 1$, one can obtain an approximate distribution for $G$. Transforming to $\Neff$ gives
\begin{eqnarray}
\label{eq:PofNeff}
P(\Neff){\rm d}\Neff = \mathcal{N}\exp\left[-\frac{4}{3}\frac{(\Neff-\avg{\Neff})^2}{\avg{\Neff}}\right]{\rm d}\Neff
\end{eqnarray}
where $\mathcal{N}$ is a normalisation factor which approaches $2/(3\pi\avg{\Neff})^{1/2}$ as $\Neff\to\infty$.  Note that in this limit one obtains by inspection $\Var(\Neff)/\avg{\Neff}=3/8$, which holds for all $\avg{\Neff}$ in a Fabry-Perot cavity, highlighting the difference between $\Neff$ and the total number of atoms in a small volume around the cavity.  We have used the full numerical distribution for analysing our results with $\avg{\Neff}=1.24(5)$. However we note that even in this regime the approximation (\ref{eq:PofNeff}) predicts $\Var(k)/\avg{k}=1.081(6)$ for our fluorescence measurements, which is still in agreement with our observations.


\section{Acknowledgments}
We thank J.~Dyne for technical expertise, and S.~Barrett for an introduction to quantum jump theory and simulations.  This work was funded by EPSRC and the Royal Society, and EU programmes HIP, AQUTE, and CHIMONO.

\section{Contributions}
J.G., M.T., J.K., and A.R. built the apparatus under the planning and supervision of E.A.H.  J.G., M.T., and A.R. took the data and J.G., M.T, and E.A.H. analysed the data.  J.G. and E.A.H. prepared the manuscript and all authors contributed to the editing.


\end{document}